\documentclass[12pt,]{article}
\usepackage{amsmath, amsthm, amscd, amsfonts, amssymb, graphicx, color}
\usepackage[bookmarksnumbered, colorlinks, plainpages]{hyperref}
\usepackage{graphics,epsfig}
\oddsidemargin 10mm \numberwithin{equation}{section}
\def\be{\begin{equation}}
\def\ee{\end{equation}}
\begin{document}
\begin{center}
{{\bf { Weak Gravitational lensing from regular Bardeen black holes
}}
 \vskip 0.50 cm
  {{Hossein Ghaffarnejad}\footnote{E-mail address:
hghafarnejad@yahoo.com} and Hassan niad\footnote{E-mail address:
niad@semnan.ac.ir}}\vskip 0.1 cm \textit{Department of Physics,
Semnan University, P.O.Box 35131-19111, Iran}}
\end{center}
\vspace{0.1cm} \begin{abstract}  In this article we study weak
gravitational lensing of regular Bardeen black hole which has scalar
charge $g$ and mass $m.$ We investigate the angular position and
magnification of non-relativistic images in two cases depending on
the presence or absence of photon sphere. Defining dimensionless
charge parameter $q=\frac{g}{2m}$ we seek to disappear photon sphere
in the case of $|q|>{24\sqrt5}/{125}$ for which the space time
metric encounters strongly with naked singularities. We specify the
basic parameters of lensing in terms of scalar charge by using the
perturbative method and found that the parity of images is different
in two cases: (a) The strongly naked singularities is present in the
space time. (b) singularity of space time is weak or is eliminated
(the black hole lens).
\end{abstract}
\section{Introduction}
When the light ray passes from the vicinity of a massive object it
is deflected because of the interaction between light and
gravitational field of the massive object. So each massive object
can act like a lens and causes a phenomena called gravitational
lensing. Depending on the amount of light deflection, the
gravitational lensing is divided to weak and strong ranges. Weak
deflection limit occurs when light ray passes far away from the
photon sphere whereas strong deflection limit occurs when light ray
passes from the vicinity of the photon sphere. In the latter case it
may turn around the lens once or more and relativistic images are
formed. Light rays which pass from the inside of the photon sphere
cannot escape from the gravitational field of lens, so no images can
be created. Many articles have investigated gravitational lensing by
using analytical approach in the weak [1-7] and strong [8-17]
deflection limits. Besides, numerical methods is also used to study
the gravitational lensing of black holes and naked singularities
[18-23]. Furthermore different approaches such as Amore et. al
[24,25] and Ayer et. al [26] in the study of gravitational lensing
are presented. Gravitational lensing is a powerful observational
tool for probing black holes at the center of galaxies. In article
[27] different approaches of gravitational lensing along with
observational prospects are reviewed. Cosmological and black hole
solutions which cause to make singularities in general relativity
are important subject because of the divergence of curvature.
Regular black holes are  metric solutions which contain horizons but
without singularities. The first regular black hole (RBH) which is
introduced by Bardeen [28], has both event and Cauchy horizons, but
with a regular center. Borde showed [29-30] that the absence of the
singularity is related to topology change in de-Sitter like core.
Bardeen black hole is spherically symmetric static solution of the
Einstein field equations coupled to a nonlinear electrodynamics
source [31] containing two characteristics namely charge g and mass
m. Gravitational lensing of the regular Bardeen black hole was
studied in the strong deflection limit by Eiroa and Sendra [17].
They used Bozza method [9] to calculate deflection angle for small
values of scalar charge $|q|<q_{ph}$ where
$q_{ph}=\frac{24\sqrt5}{125}\approx0.43$ and obtained the positions
and magnifications of the relativistic images. They also applied the
results to a supermassive black
hole located at the center of assumed Galaxy.\\
In this paper we have studied gravitational lensing of Bardeen
metric in weak deflection limit in cases where the lens treats as
black hole and naked singularities. We use perturbation approach
presented by Keeton et. al [5] and find positions of primary and
secondary non-relativistic images.  We should point that all image
angular positions are re-scaled to angular Einstein radius
($\vartheta_{E}$) throughout this paper.
Organization of the paper is given as follows.\\
In section 2 we define the Bardeen black hole metric. We specify
Taylor series expansion of the elliptical integral of deflection
angle of light ray in terms of the inverse of dimensionless impact
parameter $u=\frac{b}{2m}=\frac{1}{2m}\left|\frac{L}{E}\right|>1$
where $m$, $L$ and $E$ are the mass of gravitational lens, the
constant angular momentum and the energy of light ray respectively.
We also identify the locations of the horizons and the photon sphere
of the Bardeen black hole (lens). In section 3 we use
Virbhadra-Ellis lens equation [21] and find its Taylor series
expansion in terms of $u$. Then we determine the non-relativistic
image positions. In section 4 we obtain the magnification of the
images by using Taylor series expansion of the magnification
equation. Total and weighted-centroid magnifications are also
evaluated. In section 5 we discuss results of the our work.
\section{Bardeen black holes and deflection angle}
 Regular black holes, are the solutions of the gravity equations for which an event horizon exist,
  although there is no singularity. Usually they are supported
  by nonlinear electromagnetic fields
  and so they have at least two characteristics namely mass $m$ and charge $g$ [28,32]. They are good candidates
  for the supper-massive Galactic black holes treated as particle accelerators [33].  The first RBH metric was introduced
  by Bardeen  as follows
\be ds^2=-A(r)dt^2+B(r)dr^2+C(r)(d\theta^2+\sin^2\theta
d\varphi^2)\label{2.1}\ee
 where
\be A(r)=1-\frac{2m r^2}{\big(r^2+g^2\big)^{3/2}} \;,\;
B(r)=\frac{1}{A(r)}  \;,\; C(r)=r^2\label{2.2}\ee It can be interpreted as a magnetic solution of the
Einstein equations coupled to nonlinear electrodynamics [31]. Its
Ricci and Kretchmann scalars are calculated as
\be R^\mu_\mu={\frac {6m{g}^{2} \left( 4\,{g}^{2}-{r}^{2} \right) }{
\left( {r}^{2 }+{g}^{2} \right) ^{7/2}}}\label{2.3} \ee and
\be R_{\mu\nu\eta\lambda}R^{\mu\nu\eta\lambda}={\frac {{12m}^{2}
\left( 4\,{r}^{8}-12\,{g}^{2}{r}^{6}+47\,{g}^{4}{r
}^{4}-4\,{g}^{6}{r}^{2}+8\,{g}^{8} \right) }{ \left(
{r}^{2}+{g}^{2}
 \right) ^{7}}}\label{2.4}
\ee which have regular values in all points of the space-time for
$g\neq0.$ The above black hole metric reduces to de-Sitter like
space-time in small $r$. Defining a dimensionless parameter called
scalar charge as \be q=\frac{g}{2m}\label{2.5} \ee one can obtain
that the Bardeen black hole reduces to Schwarzschild space-time for
$q=0$ and also for in regions with large $r$ (in the case of
$q\neq0$). Its horizons disappear for $|q|>q_h$ where
$q_h=\frac{2\sqrt3}{9}\approx0.38$ and its photon sphere also
disappears for $|q|<q_{ph}$
($q_{ph}=\frac{24\sqrt5}{125}\approx0.43$), so that we have three
cases in study of gravitational lensing: (i) Regular black hole
(RBH) in case $|q|<q_h.$ (ii) weakly naked singularity (WNS) in case
$q_{h}<|q|<q_{ph}.$ (iii) strongly naked singularity (SNS) in case
$|q|>q_{ph}$. \\
It is convenient to use dimensionless elements of
the metric as \be
dS^2=(2m)^{-2}ds^2=-A(x)dT^2+B(x)dx^2+C(x)d\Omega^2\label{2.6}\ee
where
 we define
 \be x=\frac{r}{2m}~,~ T=\frac{t}{2m}\label{2.7}\ee and \be A(x)=B(x)^{-1}=1-\frac{x^2}{\big(x^2+q^2\big)^{3/2}} \;,\;
C(x)=x^2.\label{2.8}\ee We can determine  locations of the apparent
and event horizons of the Bardeen black hole $x_h$ by solving the
equation $A(x_h)=0$ as \be
x_h^6+(3q^2-1)x_h^4+3q^4x_h^2+q^6=0\label{2.9}~.\ee Diagram of the
above equation is plotted against $q$  in figure 1 (dotted line). \\
The photon sphere radius $x_{ps}$ is given by the largest positive
solution of the following equation [34] (see also Eq. 29 in Ref.
[27]). \be
\frac{C'(x_{ps})}{C(x_{ps})}=\frac{A'(x_{ps})}{A(x_{ps})}\label{2.10}
\ee
 where the prime denotes differentiation with respect to $x$.
Inserting (2.8) into (2.10), this equation leads to \be
4(x_{ps}^2+q^2)^5-9x_{ps}^8=0~.\label{2.11}\ee Diagram of the above
equation is
plotted against $q$ in figure 1 (solid line).\\
Deflection angle of light ray coming from infinity is specified by
solving the geodesic equation as \be \alpha(r_0)=2\Delta\phi(r_0)
-\pi\label{2.12}\ee where $r_0> r_{ps}$ is closest approach distance
of the light ray from center of the Bardeen black hole. Also we have
[27]:
\be\Delta\phi(r_0)=b\int_{r_{0}}^{\infty}\Bigg(\frac{B(r)}{C(r)}\Bigg)^{1/2}
\Bigg(\frac{C(r)}{A(r)}-b^2\Bigg)^{-1/2}dr \label{2.13}\ee in which
impact parameter \textit{b} is defined as \be
b=\sqrt{\frac{C(r_{0})}{A(r_{0})}}~.\label{2.14}\ee Applying (2.7)
and (2.8), the equations defined by (2.13) and (2.14) lead to the
following forms respectively \be
\Delta\phi(x_0)=\int_{x_{0}}^{\infty}\frac{x_0
dx}{x\sqrt{x^2\bigg(1-\frac{x_0^2}{\big(x_0^2+q^2\big)^{3/2}}\bigg)-x_0^2\bigg(1-\frac{x^2}{\big(x^2+q^2\big)^{3/2}}\bigg)}}
\label{2.15}\ee and
\be\frac{b}{2m}=\frac{x_0}{\sqrt{1-x_0^2/(x_0^2+q^2)^{3/2}}}~.\label{2.16}\ee
Using the transformation $z=\frac{x_0}{x},$ the equation (2.15) can
be rewritten as follows. \be
\Delta\phi(x_0)=\int_{0}^{1}\frac{dz}{\sqrt{1-\frac{x_0^2}{\big(x_0^2+q^2\big)^{3/2}}-z^2+\frac{x_0^2z^3}{\Big(x_0^2+q^2z^2\Big)^{3/2}}}}
\label{2.17}\ee which diverges to infinity at $z=1$. If we want to
evaluate (2.17) in limits of weak gravitational lensing where
$x_0>>1$, we must be find its Taylor series expansion about
$\frac{1}{x_0}$ and integrate it term by term as follows. \be
\alpha(x_0)=\frac{2}{x_0}+\Big(\frac{15\pi}{16}-1\Big)\frac{1}{x^2_0}+\Big(\frac{61}{12}-\frac{15\pi}{16}-4q^2\Big)\frac{1}{x_0^3}+
$$$$\bigg(\frac{3465\pi}{1024}-\frac{65}{8}+\Big(\frac{15}{2}-\frac{315\pi}{64}\Big)q^2\bigg)\frac{1}{x^4_0}+
$$$$\bigg(\frac{7783}{320}-\frac{3465\pi}{512}+\big(\frac{90\pi}{8}-\frac{195}{4}\big)q^2+6q^4\bigg)\frac{1}{x^5_0}+O\bigg(\frac{1}{x_0^6}\bigg)
~.\label{2.18}\ee The above convergent series expansion is described
in terms of closest distance $x_0>1$ which it is coordinate
dependent. But we should rewrite (2.18) against coordinate invariant
expression such as impact parameter of the light ray
$b=\left|\frac{L}{E}\right|.$ This is a good candidate for our
purpose where $L$ and $E$ are the constants of angular momentum and
energy of the light ray respectively [35]. This constant like the
black hole mass $m$ and charge $g$ is invariant so the above
expansion series described in terms of $b$ is coordinate-free.
Furthermore we should note that in the weak gravitational lensing
approach we must set $m/b<<1$ which is equivalent to $x_0>>1.$
Therefore we need Taylor series expansion of the function $x_0(b)$
obtained from (2.16) as follow. \be
\frac{1}{x_0}=\frac{2m}{b}+\frac{1}{2}\Big(\frac{2m}{b}\Big)^2+
\frac{5}{8}\Big(\frac{2m}{b}\Big)^3+(2-\frac{3}{2}q^2)\Big(\frac{2m}{b}\Big)^4
+\Big(\frac{231}{64}-\frac{21}{4}q^2\Big)\Big(\frac{2m}{b}\Big)^5+\cdots
.\label{2.19}\ee
 Substituting (2.19) the equation (2.18) become
\be \alpha_q(u>1)=\frac{A_1}{u}+\frac{A_2}{u^2}+\frac{A_3}{u^3}+\frac{A_4}{u^4}+\frac{A_5}{u^5}+\cdots\label{2.20}\ee
where $u=\frac{b}{2m}$ is dimensionless impact parameter
 and
\be A_1=2\;\;,A_2=\frac{15\pi}{16}\;\;,A_3=\frac{4}{3}\big(4-3q^2\big)\;\;,A_4=\frac{315\pi}{1024}\big(11-16q^2\big),
$$$$A_5=\frac{2}{5}\big(56-120q^2+15q^4\big)~.
\label{2.21}\ee The Taylor series expansion (2.20) remains
convergent for $u>\sqrt[n]{A_n}$ $;n=1,2,3,\cdots$ even if we choose
large scalar charge values ($|q|>q_{ph}$) i.e. for SNS. Diagram of
the deflection angle equation (2.20) is plotted against $u$ for
different values of scalar charge $q$ in figure 2. Figure 2-a shows
that for RBH and WNS with fixed $|q|$, the deflection angle
decreases with respect to impact parameter whereas in case of SNS,
there are some values of \textit{q} for which the diagram is not
treat monotonously. We should note that the relativistic images are
formed when $\alpha\geq\frac{3\pi}{2}\approx4.71.$ We focus here on
non-relativistic images for which diagrams in figure 2 are only
valid  for $\alpha<4.71~.$ \\
 \section{Lens equation and Image positions}
We take  Virbhadra-Ellis lens equation [21] as \be
tan\mathfrak{B}=tan\vartheta-D(tan\vartheta+tan(\alpha-\vartheta))\label{3.1}\ee
where $\mathfrak{B}$ and $\vartheta$ are source and image angular
positions measured from the optical axis respectively, and D is
defined as \be D=\frac{d_{ls}}{d_{os}}\label{3.2}\ee where $d_{ls}$
and $d_{os}$ are source-lens and source-observer distance
respectively (see figure 3). One of important quantities in study of
the gravitational lensing is angular radius of Einstein rings
\be\vartheta_E=\sqrt{\frac{4GmD}{c^2 d_{ol}}}\label{3.3}\ee where
$G$, $c$, $m$ and $d_{ol}$ are Newton`s gravitational constant,
speed of light, lens mass and lens-observer distance respectively.
Now, we define re-scaled angular parameters as
\be\beta=\frac{\mathfrak{B}}{\vartheta_E}~,~\theta=\frac{\vartheta}{\vartheta_E}~.\label{3.4}\ee
According to the postulate presented by Keeton [35], solutions of
the lens equation (3.1) are assumed to be have Taylor series
expansion as
\be\theta=\theta_0+\theta_1\varepsilon+\theta_2\varepsilon^2+\cdots\label{3.5}\ee
where $\theta_0$ is expected to be the image position in the weak
deflection limit and the coefficients $\theta_1$, $\theta_2,
\cdots,$ are correction terms of the image positions which should be
determined. Dimensionless parameter $|\varepsilon|<1$ is considered
to be order parameter of the perturbation expansion as
\be\varepsilon=\frac{2m}{b}=\frac{1}{u}~,~~~u>1~. \label{3.6}\ee
Inserting (2.20), (2.21), (3.5) and (3.6) the equation (3.1) reduces
to the following coefficients \be
\theta_0=\frac{1}{2}\Big(\beta+\sqrt{\beta^2+4}\Big)\label{3.7},\ee
\be\theta_1={\frac {15\pi( {\beta}^{2}-\beta\sqrt {{\beta}^{2}+4}+4
 ) }{64({\beta}^{2}+4)}}
\label{3.8}\ee
and
\be \theta_2={\frac {1}{6144 \left( {\beta}^{2}+4
\right) ^ {2}}}\bigg[ \left( -2048{D}^{2}-675{\pi
}^{2}-6144{q}^{2}+12288
 \right)  \left( {\beta}^{2}+4 \right) ^{5/2}+$$$$3\beta \left( 2048
{D}^{2}+225{\pi }^{2}+2048{q}^{2}-4096 \right)  \left( {\beta}^{2}
+4 \right) ^{2}+1350 \sqrt {{\beta}^{2}+4}{\pi }^{2}+$$$$ \left(
28672{D}^{2}+1350{\pi }^{2}+12288{q}^{2} -24576D-24576 \right)
\left( {\beta}^{2}+4 \right) ^{3/2}\bigg]~. \label{3.9}\ee Applying
the above coefficients, the equation (3.5), up to terms in order
$\varepsilon^3$ become \be\theta(\beta;q)\approx{\frac {1}{24576
\left( {\beta}^{2}+4 \right) ^{2}}}\bigg[\big[ \left( -675{\pi
}^{2}+24064 \right) {\beta}^{4}-2880\pi {\beta}^{3}+$$$$\left(
-4050{\pi }^{2}+162816 \right) {\beta}^{2}-11520\pi \beta-4050{\pi
}^{2}+266240 \big] \sqrt {{\beta}^{2}+ 4}+$$$$\left( 675{\pi
}^{2}+1536 \right) {\beta}^{5}+2880\pi {\beta }^{4}+ \left( 5400{\pi
}^{2}+12288 \right) {\beta}^{3}+23040\pi {\beta}^{2}+
$$$$\left( 10800{\pi }^{2}+24576 \right)
\beta+46080\pi\bigg]-{\frac { \left( \left( {\beta}^{2}+2 \right)
\sqrt {{\beta}^{2}+4}- \beta \left( {\beta}^{2}+4 \right)  \right)
{q}^{2}}{4{\beta}^{2}+ 16}} \label{3.10}\ee where we set $D=0.5$ and
$\varepsilon=0.5$ (i.e. $q=m$). Diagram of the image angular
position (3.10) is plotted against $\beta$ in figure 4 by using
different values of the scalar charge $q$. In figure 4-a we can see
that for $q<q_{ph}$, image angular positions have always positive
values while will be take negative values for $q\geq q_{ph}$ (see
figure 4-b). This means that in case the SNS lens we encounter with
image parity transition. It should be noted that the primary
(secondary) images $\theta^+ \,(\theta^-)$ correspond to the region
$\beta>0$ ($\beta<0$), namely they are formed in opposite side of
the lens such that $\theta^-(\beta)=\theta^+(-\beta)$. When we
choose $|q|<q_{ph}$ ($|q|>q_{ph}$) the images are formed in the
presence (absence) of photon sphere of the Bardeen black hole.
\\ The radius of Einstein rings
$\theta_E=|\theta_q(0)|$ are determined on the vertical axes of the
figure 4 with $\beta=0.$ Corresponding equation is given in terms
of the parameters $D,$ $q$ and $\varepsilon$ as follows
\be\theta_E=\left|1+{\frac {15\pi \varepsilon}{64}}+ \left( {\frac
{20{D}^{2}}{12}}-2 D-{\frac {675{\pi }^{2}}{8192}}-{q}^{2}+2
\right) {\varepsilon}^{2 }
 +\cdots \right|~. \label{3.11}\ee Radius of Einstein rings are given on vertical axis of figure 4 where the curves are cross
 with it ($\beta=0$). Radius of rings
takes smaller values by increasing scalar charge value for
$|q|<q_{ph}$ monotonically but not in case $|q|>q_{ph}.$ In the next
section we study magnifications of the determined images.
\section{Magnifications}
It is well known that the gravitational lensing conserves surface
brightness (because of Liouville`s theorem), but it changes the
apparent solid angle of the source. The magnification of an image is
defined by the ratio between the solid angles of the image and the
source. It is evaluated by \be\mu=\left|\mu_t
\,\mu_r\right|\label{4.1}\ee in which the tangential and radial
magnification are given by
$\mu_t=(\frac{\sin\beta}{\sin\theta})^{-1}$ and
$\mu_r=(\frac{d\beta}{d\theta})^{-1}$ respectively. Now, we need to
obtain expansion series form of the function $\mu$ against
$\varepsilon$ as follows
\be\mu=\mu_0+\mu_1\varepsilon+\mu_2\varepsilon^2+\cdots\label{4.2}\ee
where \be\mu_0={\frac { \left( {\beta}^{2}+2 \right) \sqrt
{{\beta}^{2}+4}+\beta\left( {\beta}^{2}+4 \right) }{2\beta \left(
{\beta}^{2}+4 \right) }}\label{4.3},\ee \be\mu_1=-{\frac {15\pi }{32
\left( {\beta}^{2}+4 \right) ^{3/2}}}\label{4.4}\ee and
\be\mu_2={\frac {1}{3072\beta \left( { \beta}^{2}+4 \right)
^{3}}}\bigg[-\sqrt {{\beta}^ {2}+4} \big(
2048{D}^{2}{\beta}^{4}+61440{D}^{2}{\beta}^{2}+2025 {\pi
}^{2}{\beta}^{2}+$$$$18432{\beta}^{2}{q}^{2}-36864 \left( D
 \right) {\beta}^{2}+180224{D}^{2}+8100{\pi }^{2}-36864{\beta}^{
2}+61440{q}^{2}-122880D$$$$-122880 \big)+\beta \left( {\beta}^{2}+4
\right)  \left( 16384{D}^{2}+ 2025{\pi
}^{2}+6144{q}^{2}-12288D-12288 \right) \bigg]~. \label{4.5}\ee The
above coefficients have positive parity $\mu^+$ because they are
related to primary images $\theta^+$. If we want to derive
magnification with negative parity $\mu^-$ obtained from secondary
images $\theta^-,$ we must replace $\beta$ in the equation (4.2)
with $-\beta$ as \be \mu^-(\beta)\equiv\mu^+(-\beta)~.\label{4.6}\ee
In cases of gravitational micro-lensing made from distant sources
for which positions of primary and secondary images are so close and
practically in-distinguishable, then total magnification $\mu_{tot}$
and magnification-weighted centroid $\mu_{cent}$ are two main
factors in study of the gravitational lensing. They are defined as
\be \mu_{tot}=|\mu^+|+|\mu^-|\label{4.7}\ee and \be
\mu_{cent}=\frac{\theta^+|\mu^+|+\theta^-|\mu^-|}{|\mu^+|-|\mu^-|}~.\label{4.8}\ee
Using (4.2), (4.3), (4.4), (4.5) and (4.6) one can obtain Taylor
series expansion of the equations (4.7) and (4.8) respectively as
follows. \be \mu_{tot}= {\frac {{\beta}^{2}+2}{\beta\sqrt
{{\beta}^{2}+4}}}+\frac{1}{4}\Big[1024 \left( {\beta}^{2}+4 \right)
\left( {\beta}^{2}+18 \right) {D} ^{2}+6144 \left( {\beta}^{2}+4
\right) \left( {q}^{2}-2D \right)$$$$+2025{\pi
}^{2}-12288{\beta}^{2}-49152 \Big]\varepsilon^2+\cdots\label{4.9}\ee
and \be \mu_{cent}={\frac{ \left( {\beta}^{2}+3 \right)
\beta}{{\beta}^{2}+2}}-{\frac {\beta \left( {\beta}^{2}+2 \right)
^{5}}{1536 \left({\beta}^{2}+4 \right) ^{8}}}\Big[-1024 \left(
{\beta}^{2}+4 \right) \left( {\beta}^{4}+9{\beta}^{2 }-2 \right)
{D}^{2}$$$$+2025{\pi }^{2}+6144{\beta}^{2} \left( {\beta}^{2}+4
\right)
 D+6144 \left( {q}^{2}-2 \right)  \left( {\beta}^{2}
+4 \right) \Big]\varepsilon^2+\cdots ~. \label{4.10}\ee Setting
$D=0.5,\varepsilon=0.5$ and applying (4.3), (4.4) and (4.5), one can
rewrite the magnification (4.2) against $\beta$ and $q$ as that \be
\mu[\theta_0(\beta);q]\approx{\frac {{\theta _{0}}^{4}}{{\theta
_{0}}^{4}-1}}-{\frac {15\pi { \theta _{0}}^{3}}{64 \left( {\theta
_{0}}^{2}+1 \right) ^{3}}}+{ \frac {1}{ 6144 \left( \theta _{0}^2
\left( {\theta _{0}}^{2}+1 \right)^2 -1
 \right) ^{5}}}\times$$$$\bigg[{\theta _{0}}^{2} \big(
6144{q}^{2}{\theta _{0}}^{6}+256{ \theta _{0}}^{8}+2025{\pi
}^{2}{\theta _{0}}^{4}+12288\,{q}^{2}{ \theta _{0}}^{4}-13824{\theta
_{0}}^{6}+$$$$6144{\theta _{0}}^{2}{q}^{ 2}-28160{\theta
_{0}}^{4}-13824{\theta _{0}}^{2}+256 \big) \bigg]+\cdots
\label{4.11}\ee where $\theta_0(\beta)$ should be evaluated from the
equation (3.7). Setting $D=0.5$, $\varepsilon=0.5$ one can derive
total magnification (4.9) and magnification-weighted centroid (4.10)
respectively as \be\mu_{tot}(\beta;q)\approx {\frac
{{\beta}^{2}+2}{\beta\sqrt {{\beta}^{2}+4}}}+\frac{1}{16}\Big[256
\left( {\beta}^{2}+4 \right) \left( {\beta}^{2}+18 \right) +6144
\left( {\beta}^{2}+4 \right) \left( {q}^{2}-1 \right)$$$$+2025{\pi
}^{2}-12288{\beta}^{2}-49152 \Big]+\cdots\label{4.12}\ee and \be
\mu_{cent}(\beta;q)\approx{\frac { \left( {\beta}^{2}+3 \right)
\beta}{{\beta}^{2}+2}}-{\frac {\beta \left( {\beta}^{2}+2 \right)
^{5}}{6144 \left({\beta}^{2}+4 \right) ^{8}}}\Big[-256 \left(
{\beta}^{2}+4 \right) \left( {\beta}^{4}+9{\beta}^{2 }-2
\right)$$$$+2025{\pi }^{2}+3072{\beta}^{2} \left( {\beta}^{2}+4
\right) +6144 \left( {q}^{2}-2 \right)  \left( {\beta}^{2} +4
\right) \Big]+\cdots~.\label{4.13}\ee We have plotted diagrams of
the equations (4.11), (4.12) and (4.13) in figures 5, 6 and 7
respectively by regarding different values of the dimensionless
charge $q.$ Figures 5 and 6 show that parity of the images obtained
in case $|q|>q_{ph}$ is exchanged with respect to ones which
obtained in case $|q|<q_{ph}.$ Behavior of magnification-weighted
centroid is different (figure 7), namely, it increases monotonically
for $|q|<q_{ph}$ but not for $|q|>q_{ph}.$ Also the magnification
(and total magnification) of the Einstein rings reduces to infinity
for different values of scalar charge $|q|.$ In case  $|q|<q_{ph}$
(i.e. black hole and weakly naked singularity), divergence rate
increase by increasing values of scalar charge but not in case
$|q|>q_{ph}$ (i.e. strongly naked singularity).
\section{Concluding
remarks} In this paper, we calculated light ray deflection angles
with respect to impact parameter for regular Bardeen black hole.
This metric contains two characteristics namely charge $g$ and mass
$m.$ The photon sphere of black hole appears (disappears) in region
of $|q|<q_{ph}$ ($|q|>q_{ph}$) where we defined $q=g/2m$. Applying
perturbation series expansion method presented by Keeton et al., we
obtained Taylor series expansion of corresponding non-relativistic
primary and secondary image positions and also magnifications (total
and centroid) as functions of the source position $\beta.$ Physical
effects of scalar charge were studied on the parity of images and
also the radius of Einstein rings. We have found that for RBH and
WNS, the deflection angle decreases, by increasing $|q|$. But there
are happened different behaviour for SNS. The non-relativistic image
positions  become closer to each other for different values of
charge parameter by increasing the source angular position $\beta.$
The magnification (and total magnification) of images diverges to
infinite value on the location of Einstein rings $\beta=0$ for all
values of the scalar charge. Image parity formed from SNS lens are
different with respect to image parity made from RBH or WNS lens.
Intensity of magnification-weighted centroid in case of SNS is also
different with respect to cases RBH or WNS by increasing $\beta$.
There is not obtained remarkable difference in the observable
quantities as deflection angle, image position and any type of
magnification in each cases of RBH and WNS lensing.
 \vskip 0.5 cm
 {\bf References}
\begin{description}
\item[1.] P. Schneider, J. Ehlers and E. E. Falco,
\textit{\textit{Gravitational lenses,}} Springer-Verlag, Berlin
(1992).
\item[2.] A. O. Petters, H. Levine and J. Wambsganss,
\textit{Singularity Theory and Gravitational Lensing,}
Boston-Birkhauser, (2001).
\item[3.] R. Epstein and I. I. Shapiro,
Phys. Rev. D 22, 2947 (1980).
\item[4.] M. Sereno, Phys. Rev. D
69, 023002 (2004).
\item[5.]C. R. Keeton and A. O. Petters, Phys.
Rev. D 72, 104006 (2005).
\item[6.]M. Sereno and F. De Luca, Phys.
Rev. D 74, 123009 (2006).
\item[7.]M. C. Werner and A. O. Petters,
Phys. Rev. D 76, 064024 (2007).
\item[8.] S. Frittelli, T. P.
Kling, and T. Newman, Phys. Rev. D 61, 064021 (2000).
\item[9.] V.
Bozza, Phys. Rev. D 66, 103001 (2002).
\item[10.] V. Bozza, Phys.
Rev. D 67, 103006 (2003).
\item [11.] V. Bozza, F. De Luca, G.
Scarpetta, and M. Sereno, Phys. Rev. D 72, 083003 (2005).
\item[12.] V. Bozza, F. De Luca, and G. Scarpetta, Phys. Rev. D 74, 063001 (2006).
\item[13.] R. Whisker, Phys. Rev. D 71, 064004
(2005).
\item[14.] E. F. Eiroa, Phys. Rev. D 71, 083010 (2005).
\item[15.] E. F. Eiroa, Phys. Rev. D 73, 043002 (2006).
\item[16.] K. Sarkar and A. Bhadra, Class. Quantum. Grav.23, 6101 (2006), gr-qc/0602087.
\item[17.] Eiroa, E.F. and Sendra. C.M., Class. Quantum Grav. 28, 085008 (2011).
\item[18.] Virbhadra, K. S., and Keeton, C. R., Phys. Rev. D, 77, 124014,
(2008).
\item[19.] Virbhadra, K.S., Int. J. Mod. Phys. A, 12, 4831, (1997).
\item[20.] Virbhadra, K.S., Phys. Rev. D, 79, 083004, (2009).
\item[21.] Virbhadra, K.S., and Ellis, G.F.R., Phys. Rev. D, 62, 084003, (2000).
\item[22.] Virbhadra, K.S., and Ellis, G.F.R., Phys. Rev. D, 65, 103004, (2002).
\item[23.] Virbhadra, K.S., Narasimha, D., and Chitre, S.M., Astron.
Astrophys., 337, (1998).
\item[24.] Amore, P., and Arceo, S. Phys.
Rev. D, 73, 083004, (2006).
\item[25.] Amore, P., Arceo, S., and
Fern´andez, F. M. Phys. Rev. D, 74, 083004, (2006).
\item[26.]Iyer, S. V., and
Petters, A.O., Gen. Relativ. Gravit., 39, 1563, (2007).
\item [27.] Bozza V. Gen, Rel. Grav. 42, 2269 (2010).
\item[28.] Bardeen J, Proc. GR5 (Tiflis, USSR) (1968).
\item [29.] Borde A. Phys. Rev. D50, 3692 (1994).
\item[30.] Borde A. Phys. Rev. D55, 7615 (1997).
\item [31.] Ayon Beato E and Garcia A, Phys. Lett.
B493, 149 (2000).
\item[32.] Ansoldi S, gr-qc/0802.0330 (2008).
\item[33.] Pradhan P., gr-qc/1402.2748v2 (2014).
\item[34.] C. M. Claudel, K. S. Virbhadra and G. F. R. Ellis,  J. Math. Phys. 42,  818 ,
(2001).
\item[35.]C. R. Keeton and A. O. Petters, Phys. Rev. D
72, 104006 (2005).
\item[36.] V. Bozza, Phys. Rev. D78, 103005 (2008).
\end{description}
\begin{figure}[ht!] \centering
\includegraphics[width=5in,height=4in]{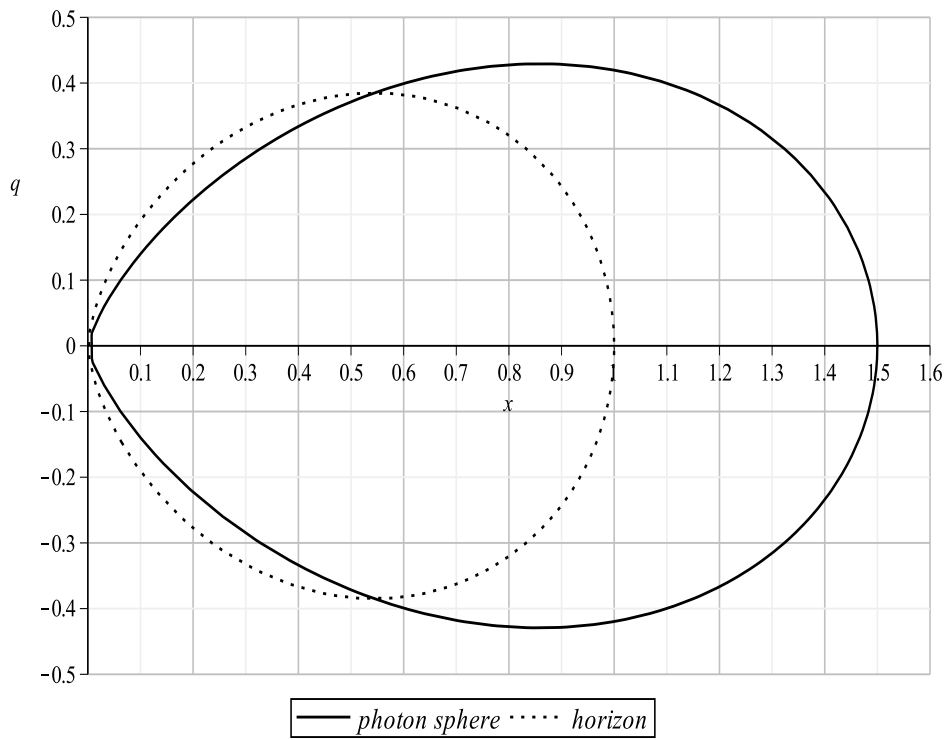}
 \caption{{\small \
Diagram of the horizon (dotted-line) and photon sphere radiuses
(solid-line) are plotted against dimensionless scalar charge $q.$}}
\end{figure}
\begin{figure}[ht!]
\centering
\includegraphics[width=5in,height=3in]{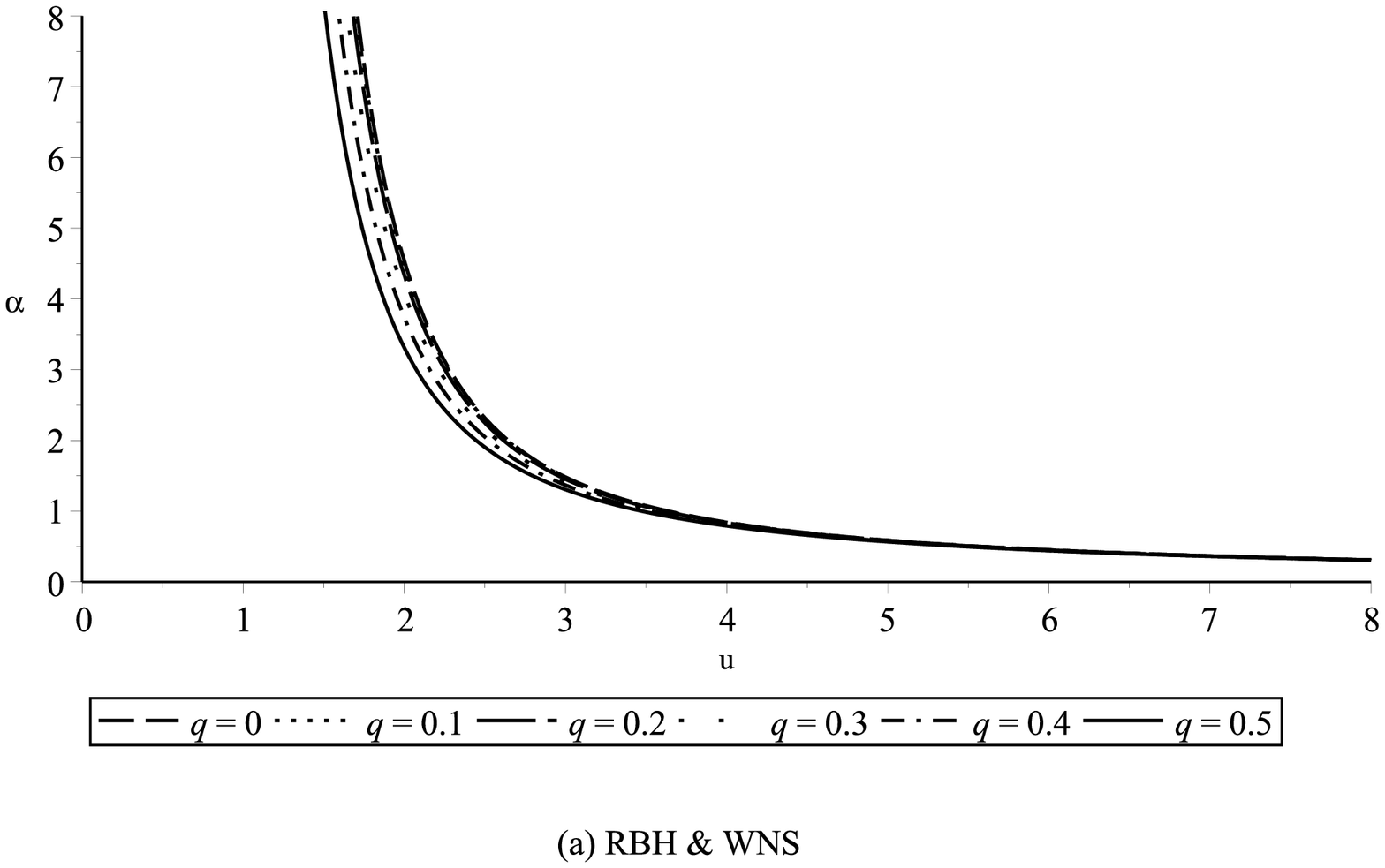}
\includegraphics[width=5in,height=3in]{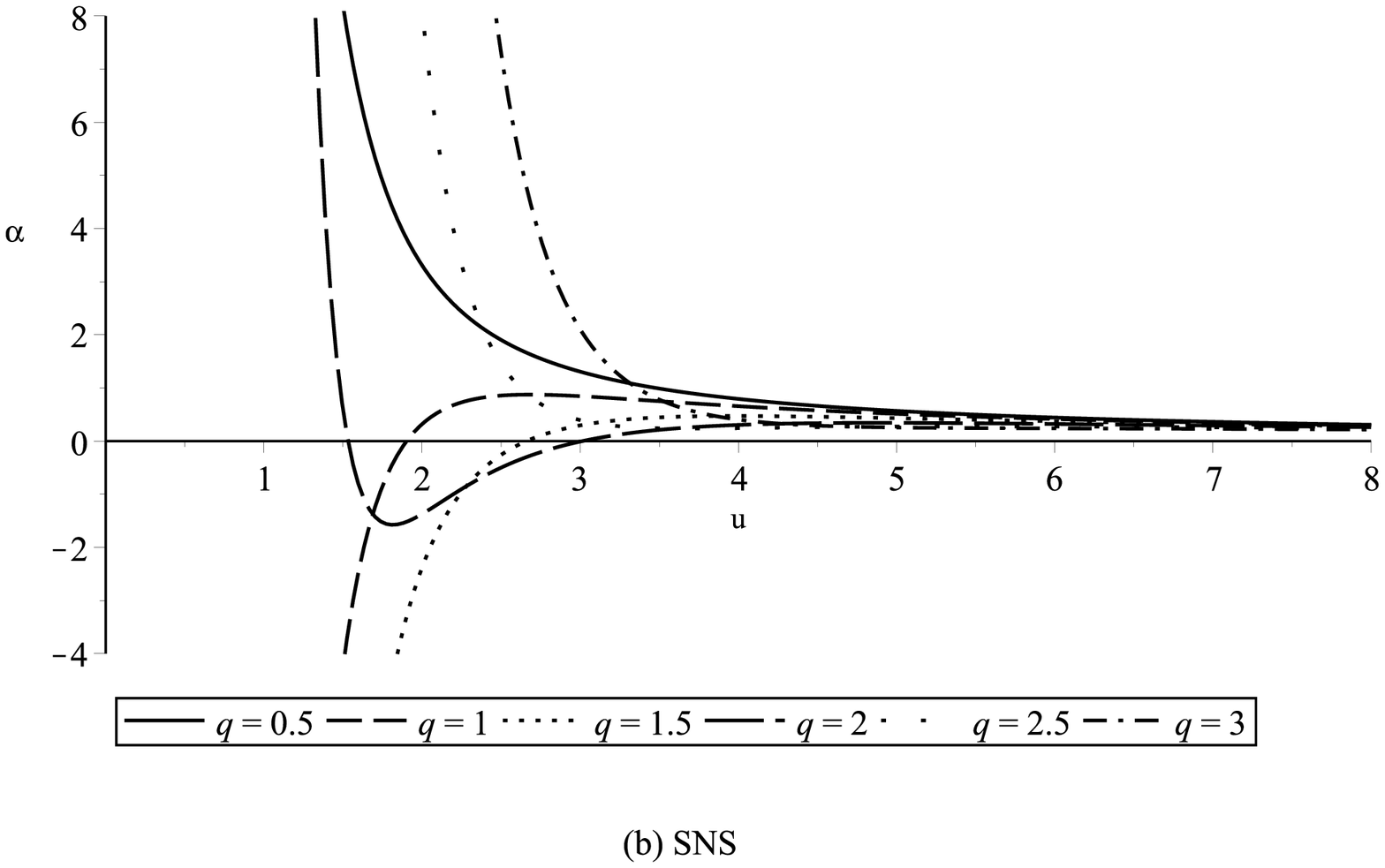}
\caption{{\small \ Diagram of deflection angle $\alpha$ is plotted
against dimensionless impact parameter $u$.}}
\end{figure}
\begin{figure}[ht!] \centering
\includegraphics[width=5in]{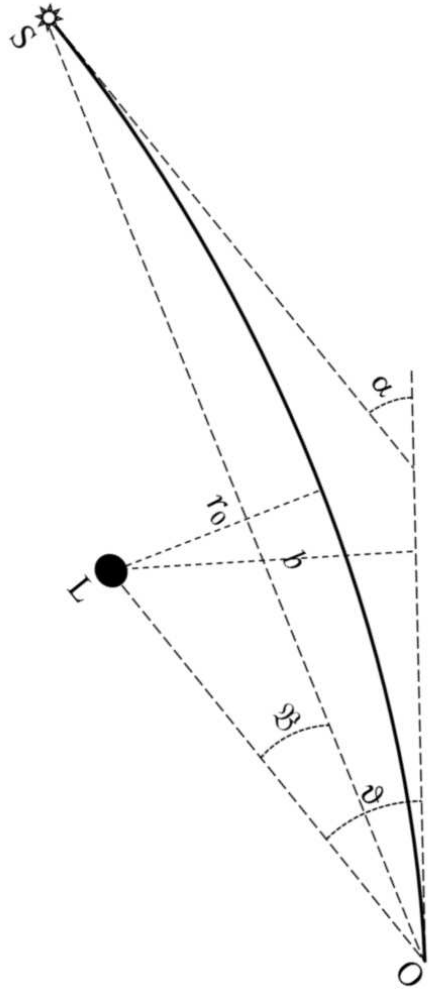}
\caption{{\small \ Gravitational lensing configuration. $\alpha$,
\textit{b} and $r_0$ are respectively deflection angle, impact
parameter and closest approach distance of the light ray.
$\mathfrak{B}$ and $\vartheta$ are the angular positions of source
and image measured from the optical axis. Position of the source,
the lens and the observer are called by S, L and O. }}
\end{figure}
\begin{figure}[ht!]
\centering
\includegraphics[width=5in,height=3in]{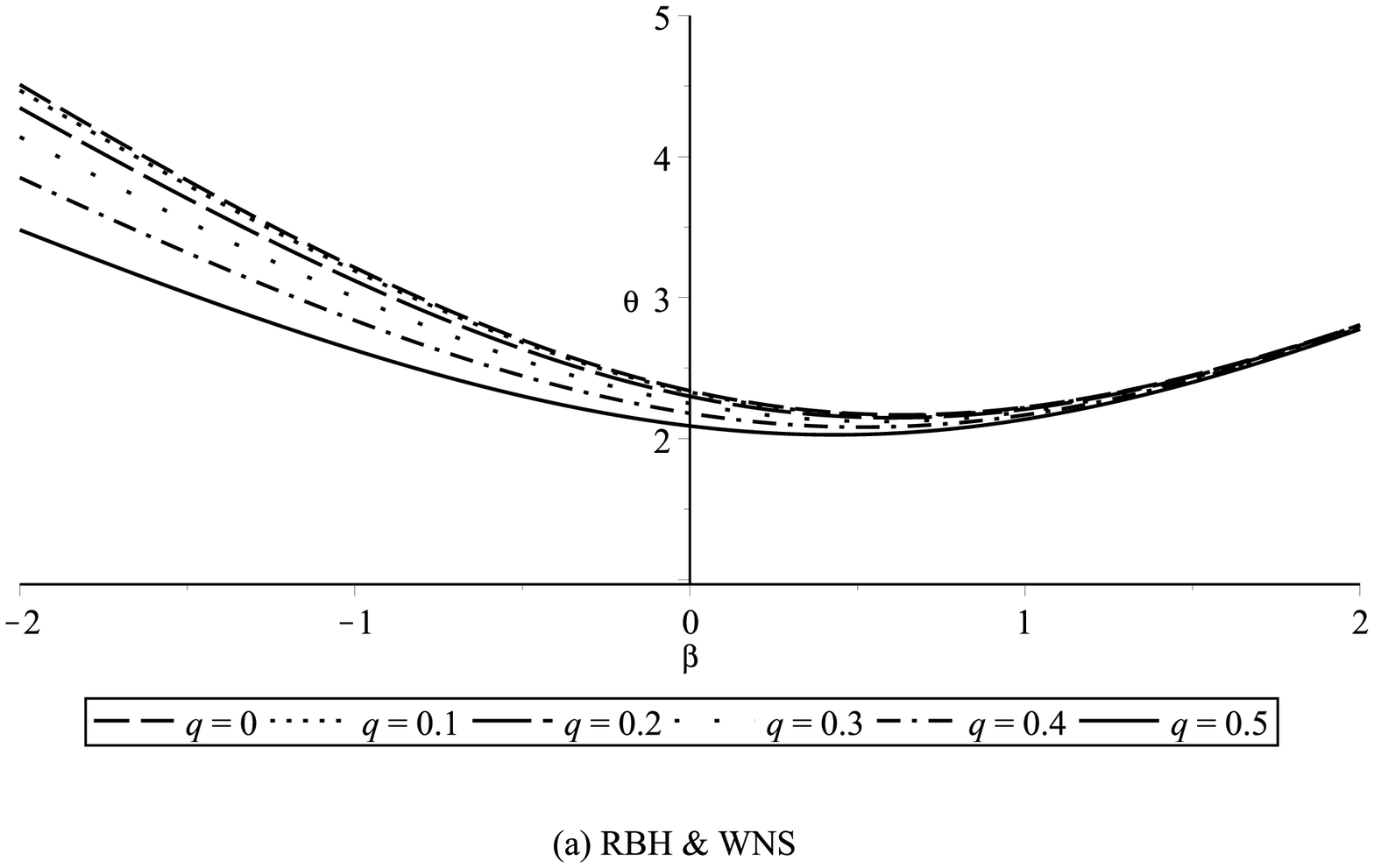}
\includegraphics[width=5in,height=3in]{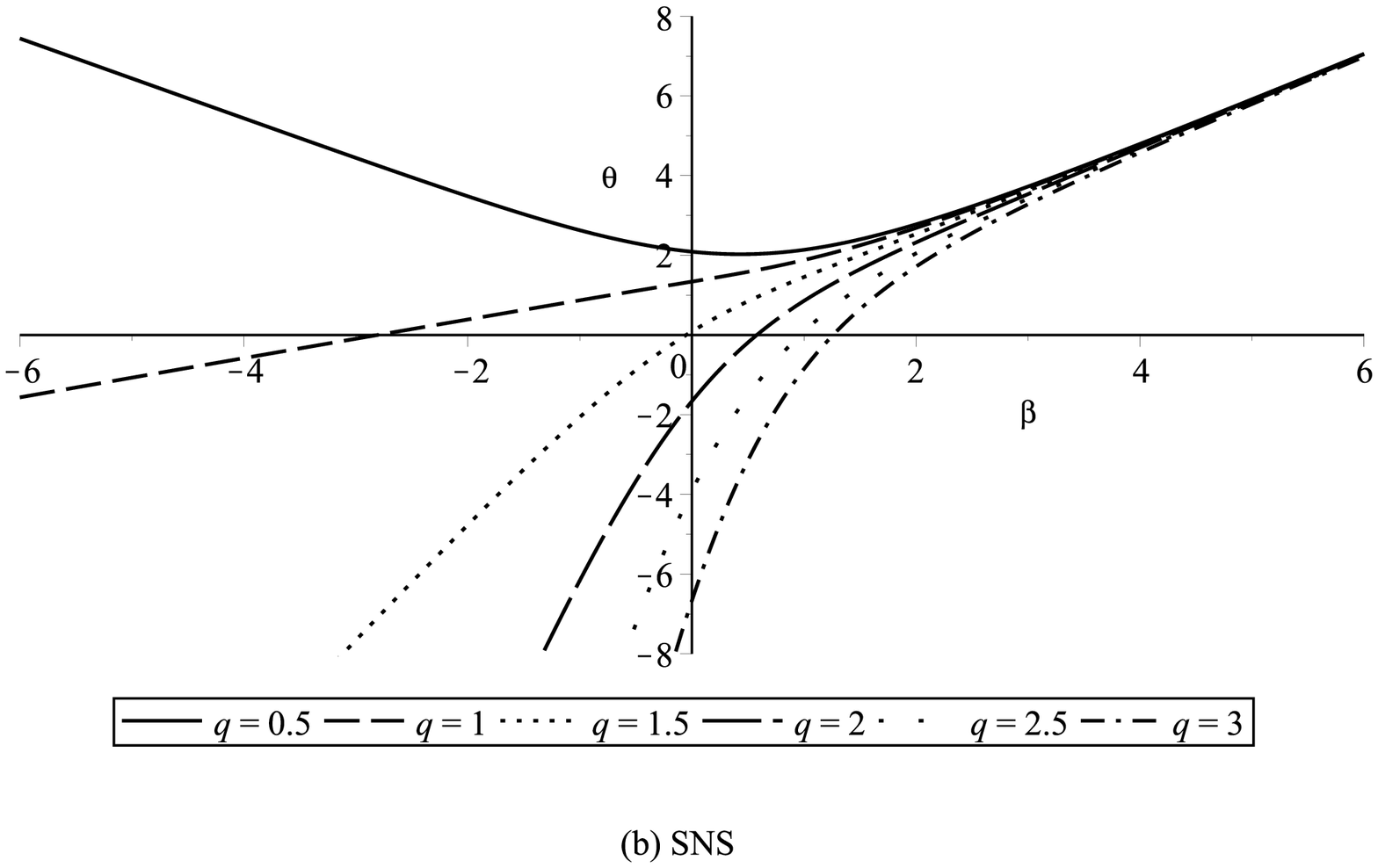}
\caption{{\small \ Diagram of image positions $\theta$ are plotted
against source position $\beta$.}} \label{fig4:figure}
\end{figure}
\begin{figure}[ht!]
\centering
\includegraphics[width=5in,height=3in]{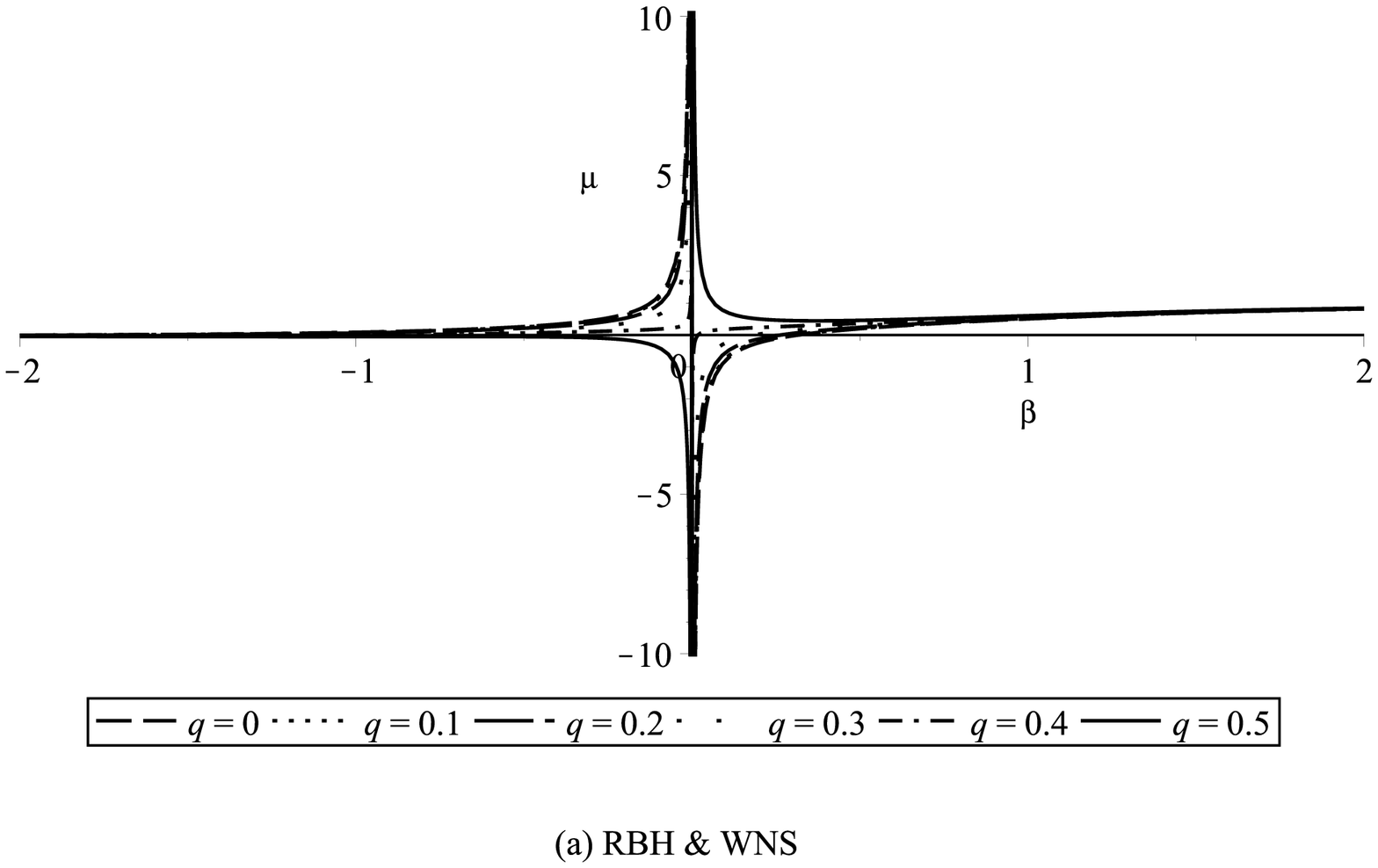}
\includegraphics[width=5in,height=3in]{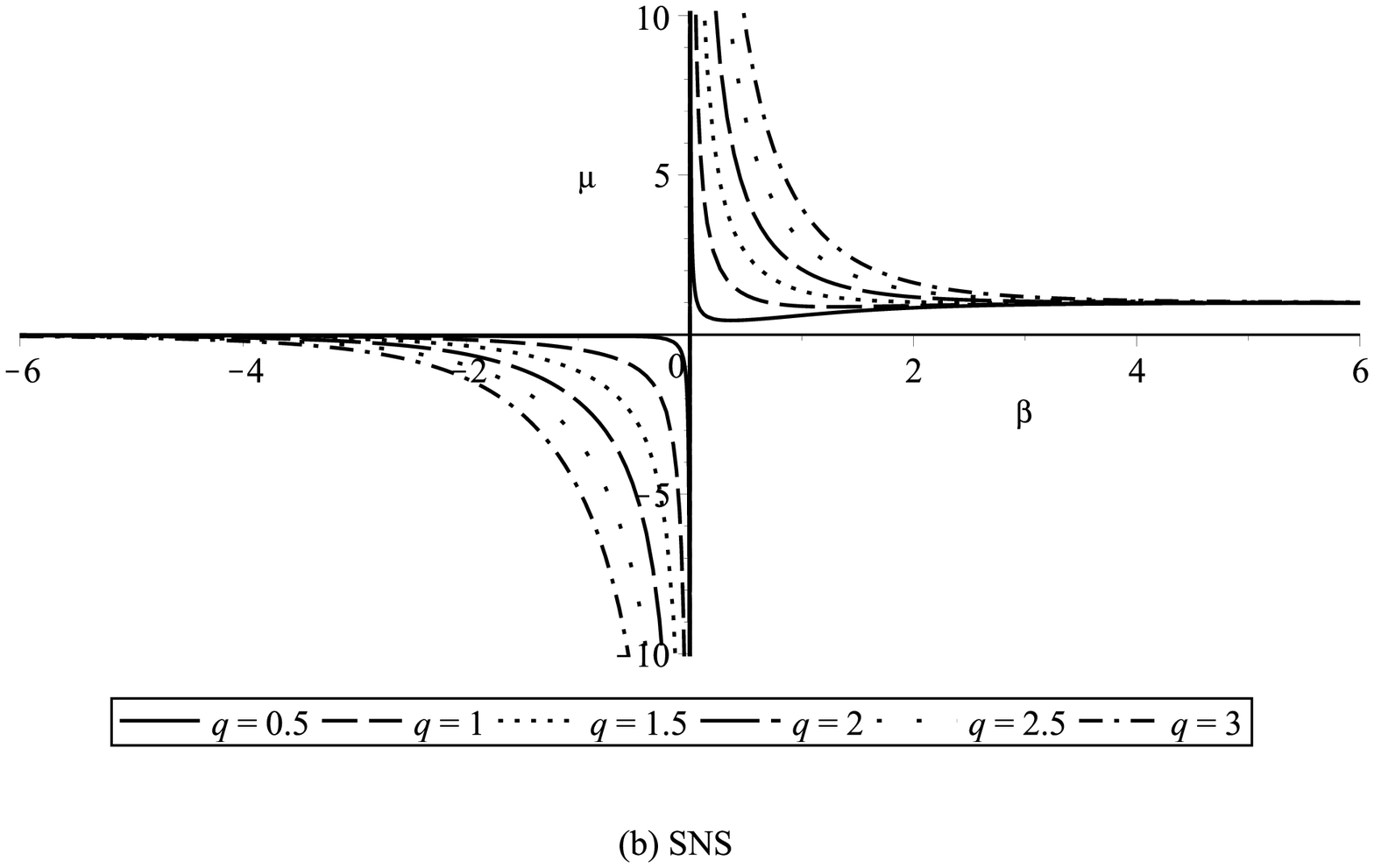}
\caption{{\small \ Diagram of magnification $\mu$ is plotted against
source position $\beta$.
 }}
\label{fig5:figure}
\end{figure}
\begin{figure}[ht!]
\centering
\includegraphics[width=5in,height=3in]{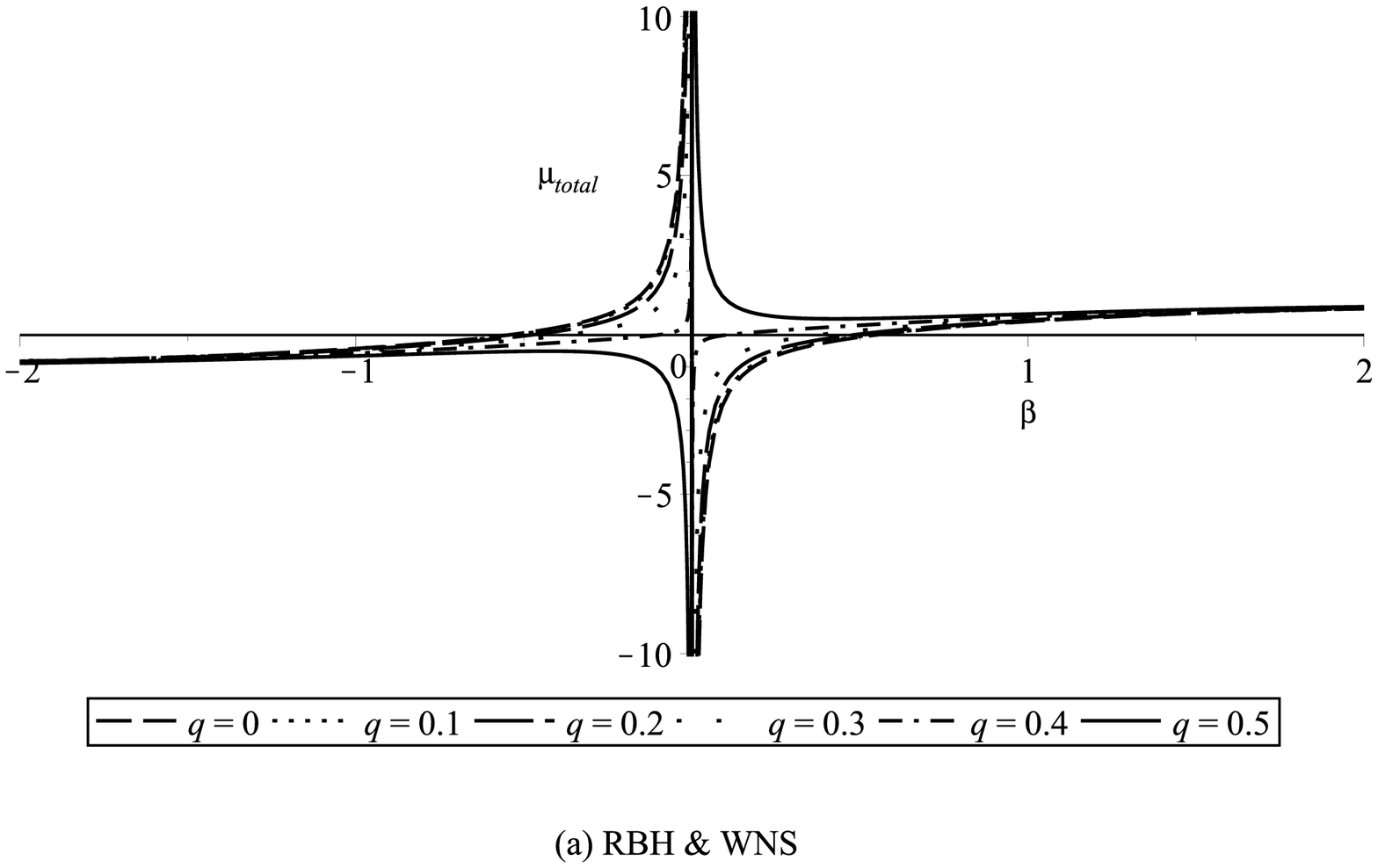}
\includegraphics[width=5in,height=3in]{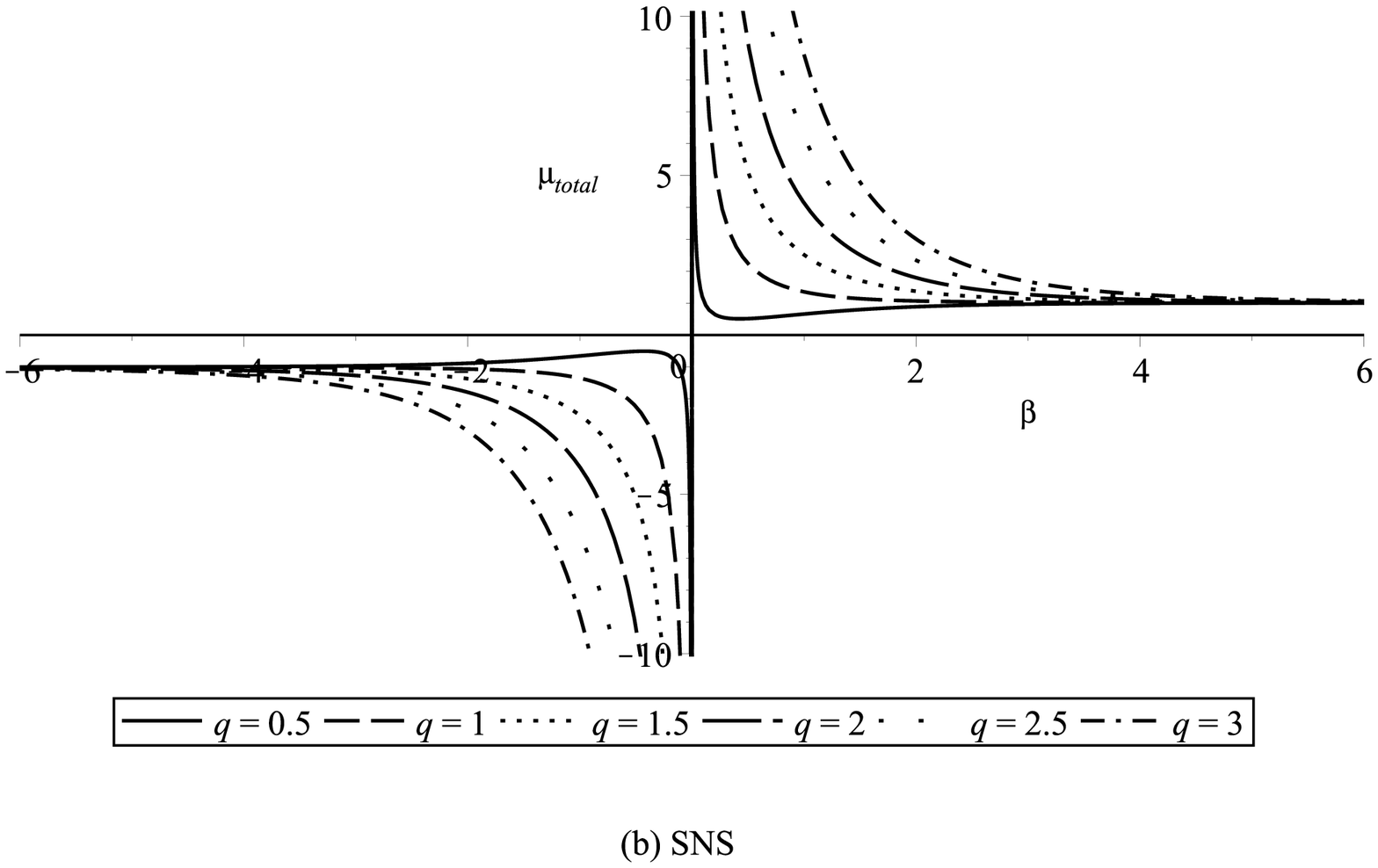}
\caption{{\small \ Diagram of total magnification $\mu_{tot}$ is
plotted against source position $\beta$.}} \label{fig6:figure}
\end{figure}
\begin{figure}[ht!]
\centering
\includegraphics[width=5in,height=3in]{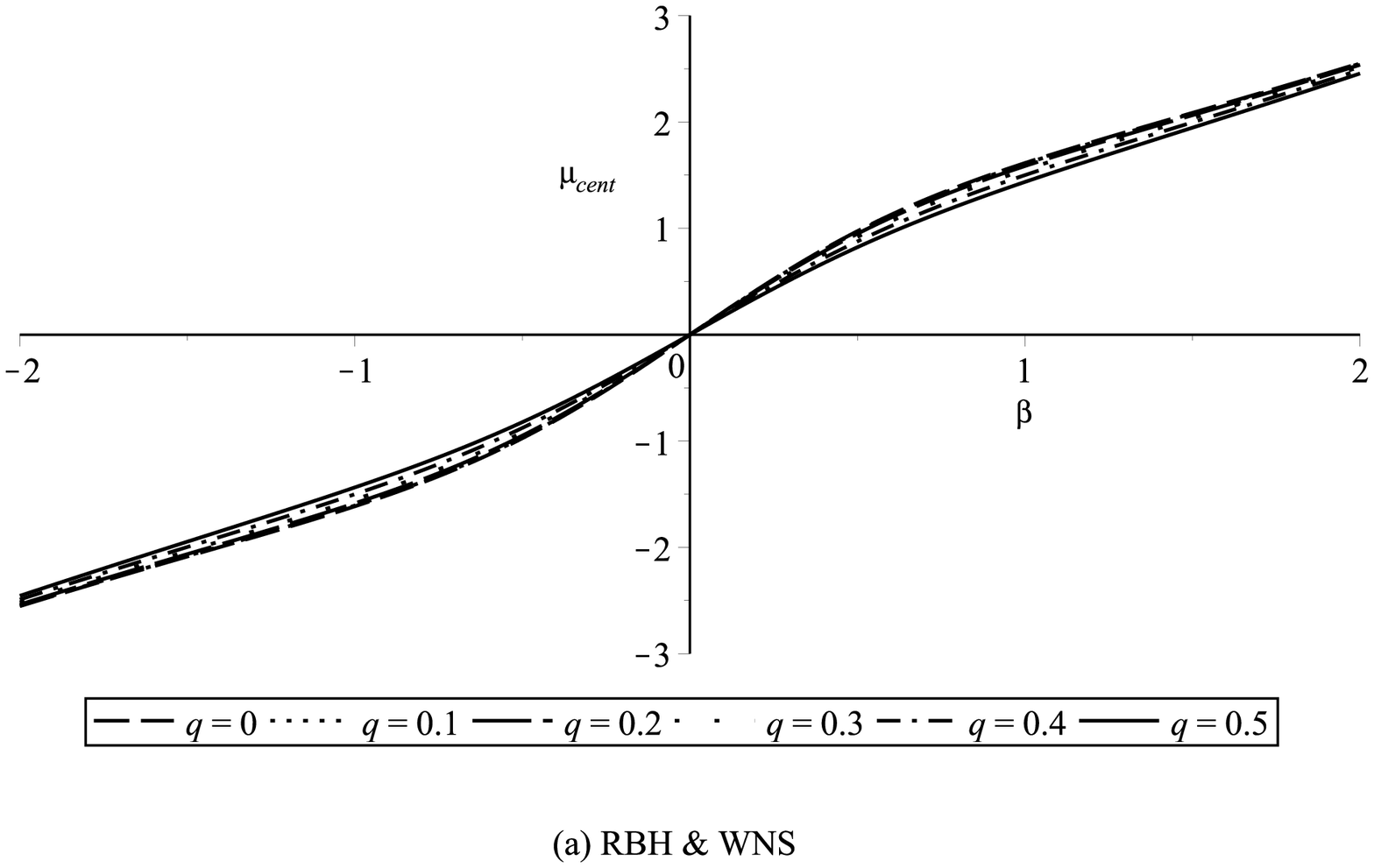}
\includegraphics[width=5in,height=3in]{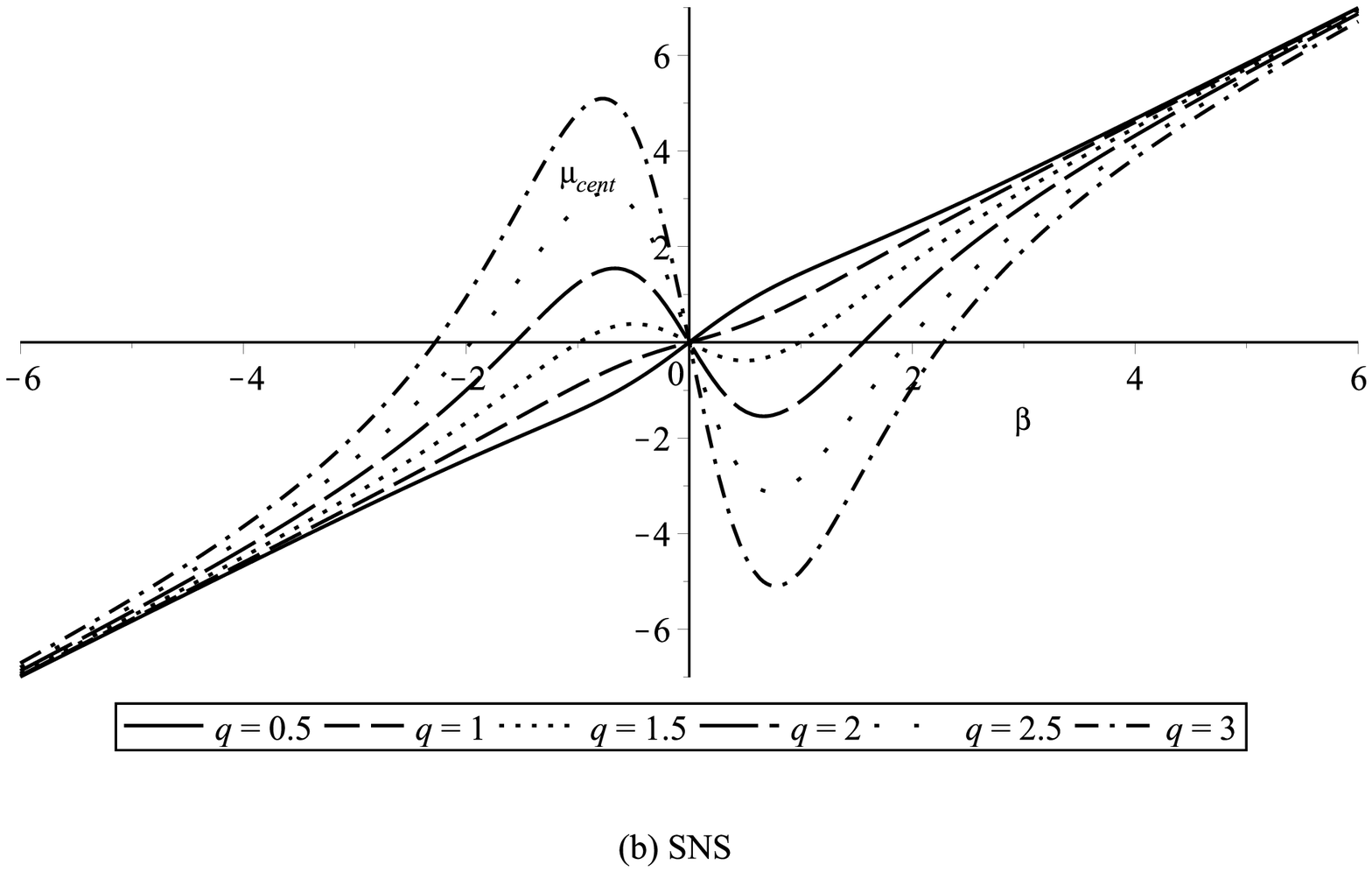}
\caption{{\small \ Diagram of the magnification-weighted centroid
$\mu_{cent}$ is plotted against source position $\beta$.}}
\label{fig7:figure}
\end{figure}
\end{document}